\documentclass[twocolumn,prb,aps,english]{revtex4-1}
\usepackage[T1]{fontenc}
\usepackage[latin9]{inputenc}
\usepackage[letterpaper]{geometry}
\usepackage{array}
\usepackage{multirow}
\usepackage{amsmath}
\usepackage{graphicx}
\usepackage{esint}

\makeatletter

\providecommand{\tabularnewline}{\\}

\@ifundefined{textcolor}{}
{%
 \definecolor{BLACK}{gray}{0}
 \definecolor{WHITE}{gray}{1}
 \definecolor{RED}{rgb}{1,0,0}
 \definecolor{GREEN}{rgb}{0,1,0}
 \definecolor{BLUE}{rgb}{0,0,1}
 \definecolor{CYAN}{cmyk}{1,0,0,0}
 \definecolor{MAGENTA}{cmyk}{0,1,0,0}
 \definecolor{YELLOW}{cmyk}{0,0,1,0}
}

\makeatother

\usepackage{babel}
\begin{document}

\title{Stochastic Optimally-Tuned Ranged-Separated Hybrid Density Functional
Theory}

\author{Daniel Neuhauser}

\email{dxn@chem.ucla.edu}

\affiliation{Department of Chemistry and Biochemistry, University of California
at Los Angeles, CA-90095 USA}

\author{Eran Rabani}

\email{eran.rabani@berkeley.edu}

\affiliation{Department of Chemistry, University of California and Materials Science
Division, Lawrence Berkeley National Laboratory, Berkeley, California
94720, USA}

\affiliation{The Sackler Center for Computational Molecular and Materials Science,
Tel Aviv University, Tel Aviv, Israel 69978}

\author{Yael Cytter}

\affiliation{Fritz Haber Center for Molecular Dynamics, Institute of Chemistry,
The Hebrew University of Jerusalem, Jerusalem 91904, Israel}

\author{Roi Baer}

\email{roi.baer@huji.ac.il}

\affiliation{Fritz Haber Center for Molecular Dynamics, Institute of Chemistry,
The Hebrew University of Jerusalem, Jerusalem 91904, Israel}

\altaffiliation{Roi Baer is on Sabbatical leave in the Department of Chemistry, University of California, Berkeley, California 94720, USA}

\begin{abstract}
We develop a stochastic formulation of the optimally-tuned range-separated
hybrid density functional theory which enables significant reduction
of the computational effort and scaling of the non-local exchange
operator at the price of introducing a controllable statistical error.
Our method is based on stochastic representations of the Coulomb convolution
integral and of the generalized Kohn-Sham density matrix. The computational
cost of the approach is similar to that of usual Kohn-Sham density
functional theory, yet it provides much more accurate description
of the quasiparticle energies for the frontier orbitals. This is illustrated
for a series of silicon nanocrystals up to sizes exceeding $3000$
electrons. Comparison with the stochastic GW many-body perturbation
technique indicates excellent agreement for the fundamental band gap
energies, good agreement for the band-edge quasiparticle excitations,
and very low statistical errors in the total energy for large systems.
The present approach has a major advantage over one-shot GW by providing
a self-consistent Hamiltonian which is central for additional post-processing,
for example in the stochastic Bethe-Salpeter approach.
\end{abstract}
\maketitle

\section{Introduction}

First-principles descriptions of quasiparticle excitations in extended
and large confined molecular systems are prerequisite for understanding,
developing and controlling molecular electronic, optoelectronic and
light-harvesting devices. In search of reliable theoretical frameworks,
it is tempting to use Kohn-Sham density functional theory (DFT),\cite{Kohn1965}
which provides accurate predictions of the structure and properties
of molecular, nanocrystal and solid state systems. However, Kohn-Sham
DFT (KS-DFT) approximations predict poorly quasiparticle excitation
energies both in confined and in extended systems,\cite{Ogut1999,Godby1998,Teale2008}
even for the frontier occupied orbital energy, for which KS-DFT is
expected to be exact.\cite{Almbladh1985,Perdew1982a,Sham1983} This
has led to the development of two main first-principles alternative
frameworks for quasiparticle excitations: many-body perturbation theory,
mainly within the so-called GW approximation\cite{Hedin1965} on top
of DFT,~\cite{Hybertsen1985,DelSole1994,Steinbeck1999,Fleszar2001,Onida2002,Rinke2005,Friedrich2006,Shishkin2007,Trevisanutto2008,Rostgaard2010,Liao2011,Blase2011a,Tamblyn2011,Samsonidze2011,Marom2012,Setten2012,Pham2013}
and generalized--KS DFT.\cite{Seidl1996,Heyd2005,Gerber2007,Brothers2008,Barone2011}

Recently, range-separated hybrid (RSH) functionals\ \cite{Savin1995a,Iikura2001,Yanai2004,Baer2005a,Livshits2007,Vydrov2006a,Chai2008}
combined with an optimally-tuned range parameter \cite{Livshits2008,Baer2010a}
were shown to very successfully predict quasiparticle band gaps, band
edge energies and excitation energies for a range of interesting \emph{small
}molecular systems, well matching both experimental results and GW
predictions.\cite{Stein2010,Kronik2012,Korzdoerfer2012,Jacquemin2014}
The key element of the range parameter tuning is the minimization
of the deviation between the highest occupied orbital energy and the
ionization energy\cite{Baer2010a,Stein2010} or the direct minimization
of the energy curvature.\cite{Stein2012}

The use of GW and the optimally-tuned RSH (OT-RSH) approaches for
describing quasiparticle excitations in \emph{extended }systems is
hampered by high computational scaling. The computational bottleneck
in GW is in the calculation of the screened potential within the Random
Phase Approximation (RPA) while in OT-RSH it is the application of
non-local exchange to each of the molecular orbitals. OT-RSH is a
self-consistent method and should therefore be compared to self-consistent
GW calculations; however, the latter are extremely expensive as the
self-energy operator must be applied to all Dyson orbitals.

Recently, we proposed a stochastic formulation limited to the $\mbox{G}_{0}\mbox{W}_{0}$
approach, where the computational complexity was reduced by combining
stochastic decomposition techniques and real-time propagation to obtain
the expectation value of the self-energy within the $\mbox{G}\mbox{W}$
approximation.\cite{Neuhauser2014a} The stochastic GW (\emph{s}GW)
was used to describe charge excitations in very large silicon nanocrystals
(NCs) with $N_{e}>3000$ ($N_{e}$ is the number of electrons), with
computational complexity scaling nearly linearly with the system size.
Similar stochastic techniques have been developed by us for DFT,\cite{Baer2013}
for embedded DFT,\cite{Neuhauser2014} and for other electronic structure
problems.\cite{Baer2012a,Neuhauser2013,Neuhauser2013a,Ge2013,Gao2014TDsDFT} 

Here we develop a stochastic formalism suitable for applying the OT-RSH
functionals for studying quasiparticle excitations in \emph{extended}
systems. The approach builds on our previous experience with the exchange
operator,\cite{Baer2012,Cytter2014,Rabani2015} but several new necessary
concepts are developed here for the first time. We start with a brief
review of the OT-RSH approach, then move on to describe the specific
elements of the stochastic approach, and finally present results.

We dedicate this paper to Prof. Ronnie Kosloff from the Hebrew University
to acknowledge his important contributions to the field of computational/theoretical
chemistry. Kosloff has been our teacher and mentor for many years
and his methods, such as the Chebyshev expansions and Fourier grids,\cite{Kosloff1988,Kosloff1994}
are used extensively in our present work as well.

\section{Optimally-Tuned Range Separated Hybrid Functionals}

For a systems of $N_{e}$ electrons in an external one-electron potential
$v_{ext}\left(\mathbf{r}\right)$ having a total spin magnetization
$s_{z}$ in the $z$ direction, the OT-RSH energy is a functional
of the spin-dependent density matrix (DM) $\rho_{\uparrow,\downarrow}\left(\mathbf{r},\mathbf{r}'\right)$
given in atomic units as:
\begin{eqnarray}
E_{RSH}^{\gamma}\left[\rho_{\uparrow},\rho_{\downarrow}\right] & = & tr\left[\rho\left(-\frac{1}{2}\hat{\mathbf{\nabla}}^{2}+v_{ext}\left(\hat{\mathbf{r}}\right)\right)\right]\\
 & + & E_{H}\left[n\right]+E_{XC}^{\gamma}\left[n\right]+K_{X}^{\gamma}\left[\rho_{\uparrow},\rho_{\downarrow}\right],\nonumber 
\end{eqnarray}
where $\gamma$ is the range-parameter, discussed below, while 
\begin{equation}
E_{H}\left[n\right]=\frac{1}{2}\iint u_{C}\left(\left|\mathbf{r}-\mathbf{r}'\right|\right)n\left(\mathbf{r}\right)n\left(\mathbf{r}'\right)d\mathbf{r}d\mathbf{r}'
\end{equation}
is the Hartree energy functional of the density $n\left(\mathbf{r}\right)=\rho\left(\mathbf{r},\mathbf{r}\right)=\sum_{\sigma=\uparrow\downarrow}\rho_{\sigma}\left(\mathbf{r},\mathbf{r}\right)$
and $u_{C}\left(r\right)=r^{-1}$ is the Coulomb potential energy.
$E_{XC}^{\gamma}\left[n\right]$ is the unknown $\gamma$--dependent
exchange-correlation energy functional which in practical applications
is approximated. The non-local exchange energy functional is given
by 
\begin{equation}
K_{X}^{\gamma}\left[\rho_{\uparrow},\rho_{\downarrow}\right]=-\frac{1}{2}\sum_{\sigma=\uparrow,\downarrow}\iint u_{C}^{\gamma}\left(\left|\mathbf{r}-\mathbf{r}'\right|\right)\left|\rho_{\sigma}\left(\mathbf{r},\mathbf{r}'\right)\right|^{2}d\mathbf{r}d\mathbf{r}',
\end{equation}
where $u_{C}^{\gamma}\left(r\right)=r^{-1}\mbox{erf}\left(\gamma r\right)$.
This choice of $u_{C}^{\gamma}\left(r\right)$ accounts for long-range
contributions to the non-local exchange energy and thus dictates a
complementary cutoff in the \emph{local} exchange-correlation energy,
$E_{XC}^{\gamma}\left[n\right]$, to avoid over-counting the exchange
energy.\cite{Savin1995,Iikura2001,Baer2010a}

When the exact $E_{XC}^{\gamma}\left[n\right]$ functional is used,
minimizing $E_{RSH}^{\gamma}\left[\rho_{\uparrow},\rho_{\downarrow}\right]$
with respect to $\rho_{\sigma}\left(\mathbf{r},\mathbf{r}'\right)$
under the constraints specified below leads to the exact ground-state
energy and electron density $n\left(\mathbf{r}\right)$. For approximate
$E_{XC}^{\gamma}\left[n\right]$ approximate estimates of these quantities
are obtained. To express the constraints we first require the spin-dependent
DM to be Hermitian and thus expressible as: 
\begin{equation}
\rho_{\sigma}\left(\mathbf{r},\mathbf{r}'\right)=\sum_{j}f_{j,\sigma}\phi_{j,\sigma}\left(\mathbf{r}\right)\phi_{j,\sigma}^{*}\left(\mathbf{r}'\right).\label{eq:DM_sigma}
\end{equation}
where $f_{j,\sigma}$ and $\phi_{j,\sigma}\left(\mathbf{r}\right)$
are its eigenvalues and orthonormal eigenfunctions. The constraints
are then given in terms of the eigenvalues $f_{j\sigma}$ as:
\begin{equation}
0\le f_{j,\sigma}\le1,
\end{equation}

\begin{equation}
\sum_{j,\sigma}f_{j,\sigma}=N_{e},
\end{equation}
\begin{equation}
\frac{1}{2}\sum_{j}\left(f_{j,\uparrow}-f_{j,\downarrow}\right)=s_{z}.
\end{equation}
The necessary conditions for a minimum of $E_{RSH}^{\gamma}\left[\rho_{\uparrow},\rho_{\downarrow}\right]$
is that $\phi_{j,\sigma}\left(\mathbf{r}\right)$ obey the generalized
KS equations:
\begin{equation}
\hat{h}_{\sigma}^{\gamma}\phi_{j,\sigma}^{\gamma}\left(\mathbf{r}\right)=\varepsilon_{j,\sigma}^{\gamma}\phi_{j,\sigma}^{\gamma}\left(\mathbf{r}\right),\label{eq:GKS-eqns}
\end{equation}
where $\varepsilon_{j,\sigma}^{\gamma}$ are the spin-dependent eigenvalues
of the generalized KS Hamiltonian ($j=1,2,\dots$ and $\sigma=\uparrow\downarrow$)
given by: 
\begin{equation}
\hat{h}_{\sigma}^{\gamma}=-\frac{1}{2}\hat{\nabla}^{2}+v_{\sigma}^{\gamma}\left(\hat{\mathbf{r}}\right)+\hat{k}_{\sigma}^{\gamma}.\label{eq:GKS-h}
\end{equation}
Note that the DM and its eigenstates minimizing the energy functional
$E_{RSH}^{\gamma}\left[\rho_{\uparrow},\rho_{\downarrow}\right]$
are themselves $\gamma$--dependent and are thus denoted by $\rho_{\sigma}^{\gamma}\left(\mathbf{r},\mathbf{r}'\right)$,
$\phi_{j,\sigma}^{\gamma}\left(\mathbf{r}\right)$; the DM eigenvalues
are not $\gamma$--dependent, as shown below. The one-electron Hamiltonian
$\hat{h}_{\sigma}^{\gamma}$ contains the kinetic energy, a local
potential in \emph{$r$}--space $v_{\sigma}^{\gamma}\left(\hat{\mathbf{r}}\right)$
and a non-local exchange operator $\hat{k}_{\sigma}^{\gamma}$. The
local $r$--space potential is further decomposed into three contributions:
\begin{equation}
v_{\sigma}^{\gamma}\left(\mathbf{r}\right)=v_{ext}\left(\mathbf{r}\right)+v_{H}\left(\mathbf{r}\right)+v_{XC,\sigma}^{\gamma}\left(\mathbf{r}\right),\label{eq:v-sigma(r)}
\end{equation}
where $v_{H}\left(\mathbf{r}\right)=\frac{\delta E_{H}\left[n\right]}{\delta n\left(\mathbf{r}\right)}=\int n\left(\mathbf{r}\right)u_{C}\left(\left|\mathbf{r}-\mathbf{r}'\right|\right)d\mathbf{r}'$
is the Hartree potential and $v_{XC,\sigma}^{\gamma}\left(\mathbf{r}\right)=\frac{\delta E_{XC}^{\gamma}\left[n\right]}{\delta\rho_{\sigma}\left(\mathbf{r},\mathbf{r}\right)}$
is the short-range exchange-correlation potential. The non-local exchange
operator $\hat{k}_{\sigma}^{\gamma}=\left.\frac{\delta K_{X}^{\gamma}}{\delta\hat{\rho}_{\sigma}}\right|_{\left[\rho_{\uparrow}^{\gamma},\rho_{\downarrow}^{\gamma}\right]}$
is expressed by its operation on a wave function $\psi_{\sigma}\left(\mathbf{r}\right)$
of the same spin as: 
\begin{eqnarray}
\hat{k}_{\sigma}^{\gamma}\psi_{\sigma}\left(\mathbf{r}\right) & = & -\int u_{C}^{\gamma}\left(\left|\mathbf{r}-\mathbf{r}'\right|\right)\rho_{\sigma}^{\gamma}\left(\mathbf{r},\mathbf{r}'\right)\psi_{\sigma}\left(\mathbf{r}'\right)d\mathbf{r}'.\label{eq:LR-X}
\end{eqnarray}

In this work we consider closed shell systems where $s_{z}=0$ and
$N_{e}=2N_{H}$ where $N_{H}$ is the number of electron pairs, i.e.,
the level number of the highest occupied orbital. In this case, as
in Hartree--Fock theory and DFT, the DM eigenvalues $f_{j\sigma}$
which minimize $E_{RSH}^{\gamma}\left[\rho_{\uparrow},\rho_{\downarrow}\right]$
are $f_{j,\sigma}=1$ if $j\le N_{H}$ and 0 otherwise.\cite{Dreizler1990}
Hence, these conditions are used \emph{a-priori} as constraints during
the minimization of $E_{RSH}^{\gamma}\left[\rho_{\uparrow},\rho_{\downarrow}\right]$.
However, for the tuning process the ensemble partial ionization of
an up-spin (or down-spin) electron needs to be considered. Thus, these
values for $f_{j,\sigma}$ are still used except for $j=N_{H}$ and
$\sigma=\uparrow$ where $f_{H,\uparrow}$ is fixed to be a\emph{
}positive fraction (i.e., the negative of the overall charge of the
system, $-c$) during the minimization of the GKS ensemble energy
$E_{RSH}^{\gamma}\left[\rho_{\uparrow},\rho_{\downarrow}\right]$
(for clarity, we abbreviate $N_{H}\equiv H$ for the frontier orbital
energy ($\varepsilon$) and occupation ($f$)). We note in passing
that tuning is often done by combining a linearity condition from
the $N+1$ electron system.\cite{Stein2009a} We leave this for future
work, and state that it can be done along the same lines as described
here for the $N$ electron system.

\begin{figure*}[t]
\begin{centering}
\includegraphics[height=6cm]{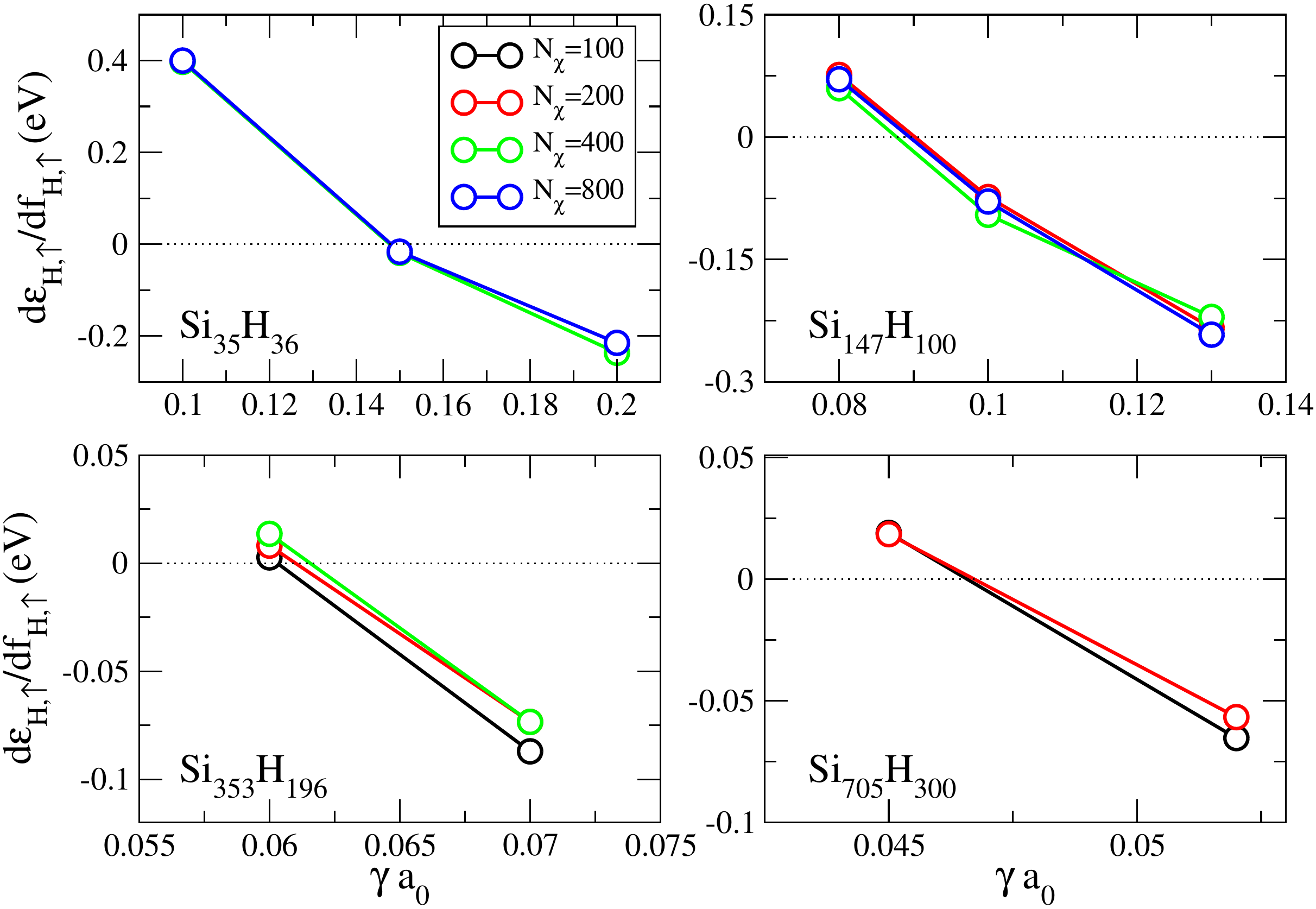}\ \ \includegraphics[height=6cm]{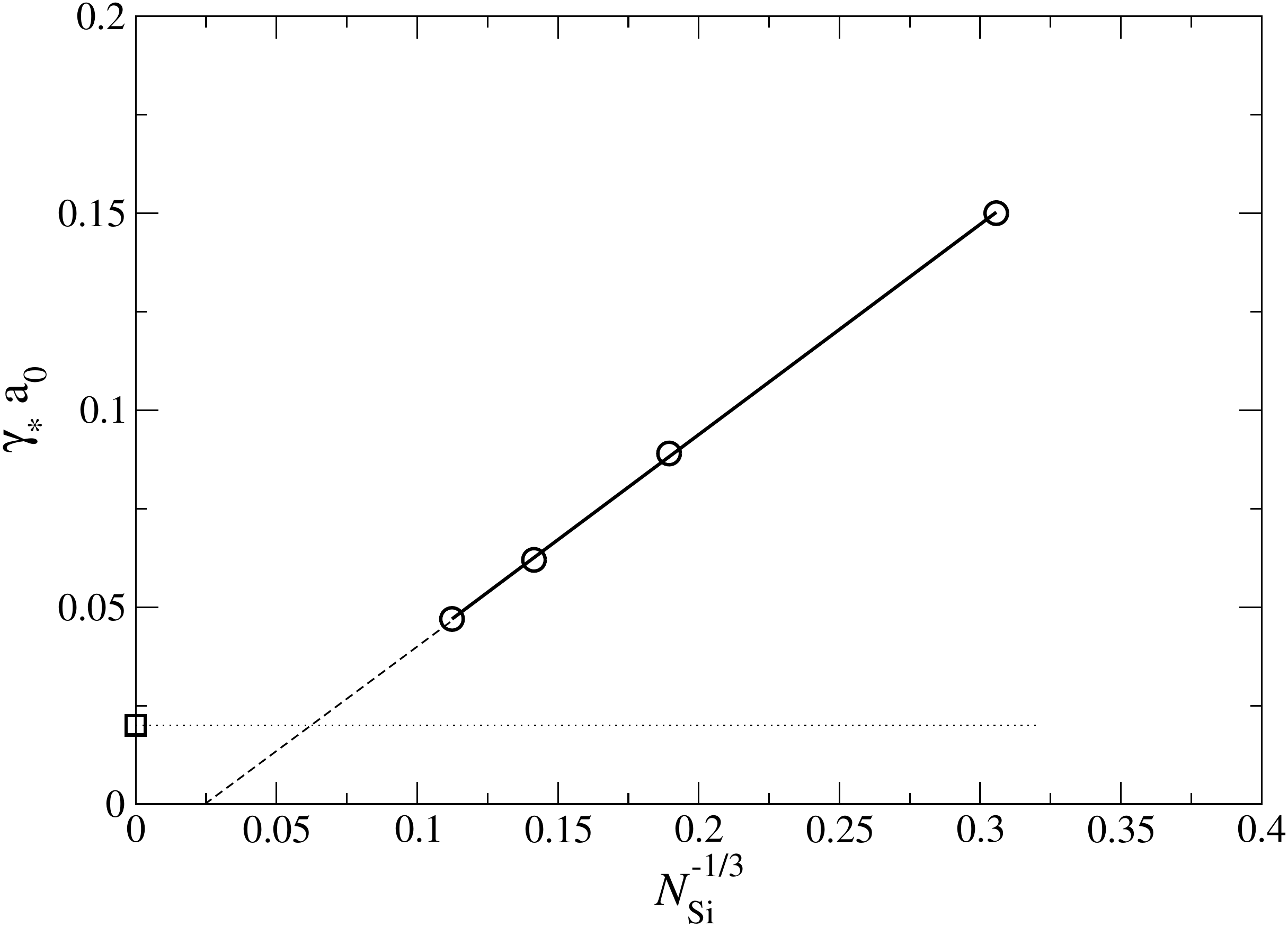}
\par\end{centering}

\protect\caption{\label{fig:gamma} Left panels: The curvature as a function of $\gamma$
for the HOMO energy, $\partial\varepsilon_{H,\uparrow}/\partial f_{H,\uparrow}$
for different silicon nanocrystals and for different number of stochastic
orbitals used to evaluate the non-local exchange. Right panel: The
optimal value of $\gamma$ determined by Eq.~\eqref{eq:de/df} for
the selected silicon nanocrystals. The results are best-fitted to
$-0.013+0.53N_{{\rm Si}}^{-1/3}$. The square is the reverse engineered
value of $\gamma$ which yields the experimental band gap of bulk
silicon Ref.\ \citenum{Eisenberg2009}. }
\end{figure*}

The optimally-tuned range-parameter $\gamma$ is determined from the
requirement that the highest occupied generalized KS orbital energy
$\varepsilon_{H,\sigma}^{\gamma}$ is independent of its occupancy
$f_{H,\sigma}$:
\begin{equation}
\frac{\partial\varepsilon_{H,\uparrow}^{\gamma}}{\partial f_{H,\uparrow}}=0.\label{eq:de/df}
\end{equation}
Through Janak's theorem~\cite{Janak1978} this equation implies that
the energy curvature $\frac{\partial^{2}E_{RHS}^{\gamma}}{\partial f_{H,\uparrow}^{2}}$
is zero. In practical terms, Eq.~\eqref{eq:de/df} is solved by a
graphical root search as shown in Fig.~\ref{fig:gamma} and discussed
below.

\section{Stochastic Formulation of the Non-Local Exchange Operator}

In real-space or plane-waves implementations the application of the
Hamiltonian $\hat{h}_{KS}$ on a single particle wave function involves
a pair of Fast Fourier Transforms (FFT) to switch the wave function
between $k$--space where the kinetic energy is applied and $r$--space
for applying the potential energy.\cite{Kosloff1983} Therefore, for
a grid of $N_{g}$ grid-points the operational cost is $10N_{g}log_{2}N_{g}$.
The KS Hamiltonian operation scales quasi linearly with system size.
The scaling is much steeper for the RSH Hamiltonian because the non-local
exchange operator $\hat{k}_{\sigma}^{\gamma}$ applies $N_{e}$ Coulomb
convolution integrals, each of which is done using an FFT of its own
thus involving $10N_{g}log_{2}N_{g}\times N_{e}$ operations. Therefore,
the GKS Hamiltonian operation which scales quasi-quadratically is
much more time consuming than the KS Hamiltonian. Our approach, described
next, reduces significantly the operation cost and even lowers the
scaling due to the reduction of $\gamma_{*}$ as the system size grows.

We first express the occupations in the DM in Eq.\ \eqref{eq:DM_sigma}
as a combination of a occupations of a closed-shell density matrix
and a remnant due to the overall charge of the molecule, $c$ (assuming
$-1\le c\le1)$; This separation reduces the stochastic error later
when the charge of the system is continuously varied, as needed for
the optimal tuning. Thus:

\begin{eqnarray}
\rho_{\sigma}\left(\mathbf{r},\mathbf{r}'\right) & = & \sum_{j}f_{j,\sigma}\phi_{j,\sigma}\left(\mathbf{r}\right)\phi_{j,\sigma}^{*}\left(\mathbf{r}'\right)\\
 & = & \sum_{j\le N_{H}}\phi_{j,\sigma}\left(\mathbf{r}\right)\phi_{j,\sigma}^{*}\left(\mathbf{r}'\right)-c\phi_{F\uparrow}\left(\mathbf{r}\right)\phi_{F\uparrow}^{*}\left(\mathbf{r}'\right)\label{eq:chargedDM}
\end{eqnarray}
where $\phi_{F\uparrow}$ is the frontier orbital being charged and
$c$ is the amount of charge. When tuning the neutral system $F=H$
is the HOMO and it is being positively charged (electrons removed
from HOMO) so $c<0$. When tuning for the anion $F=H+1$ is the LUMO
and the system being negatively charged (electrons are added to the
LUMO) so $c>0$. We assume without loss of generality that the spin
of the charge frontier orbital is up. Next we evaluate the first term
on the RHS of Eq.~\eqref{eq:chargedDM} using stochastic orbitals:
\begin{equation}
\sum_{j\le N_{H}}\phi_{j,\sigma}\left(\mathbf{r}\right)\phi_{j,\sigma}^{*}\left(\mathbf{r}'\right)=\left\langle \eta_{\sigma}\left(\mathbf{r}\right)\eta_{\sigma}^{*}\left(\mathbf{r}'\right)\right\rangle _{\xi},
\end{equation}
where $\eta_{\sigma}\left(\mathbf{r}\right)$ is a projected-stochastic
orbital described in terms of the eigenstates of $\hat{h}_{\sigma}$
(which can be alternatively obtained using a Chebyshev expansion of
the relevant projection operator\ \cite{Baer2013}):
\begin{equation}
\eta_{\sigma}\left(\mathbf{r}\right)=\sum_{j\le N_{H}}\phi_{j,\sigma}\left(\mathbf{r}\right)\left\langle \left.\phi_{j,\sigma}\right|\xi\right\rangle ,\label{eq:eta-sigma}
\end{equation}
and 
\begin{equation}
\xi\left(\mathbf{r}\right)=\pm\frac{1}{\sqrt{h^{3}}}
\end{equation}
is a stochastic orbital with a random sign ($\pm1$) at each grid-point.
(Note that application of Eq.~\eqref{eq:eta-sigma} is itself a quadratic
step however it is a ``cheap'' step as it is done only once in each
SCF iteration.) With this, Eq.\ \eqref{eq:LR-X} is rewritten as:
\begin{align*}
\hat{k}_{\sigma}^{\gamma}\psi_{\sigma}\left(\mathbf{r}\right)= & -\left\langle \eta_{\sigma}\left(\mathbf{r}\right)\int u_{C}^{\gamma}\left(\left|\mathbf{r}-\mathbf{r}'\right|\right)\eta_{\sigma}^{*}\left(\mathbf{r}'\right)\psi_{\sigma}\left(\mathbf{r}'\right)d\mathbf{r}'\right\rangle _{\xi}
\end{align*}
\begin{equation}
+c\int u_{C}^{\gamma}\left(\left|\mathbf{r}-\mathbf{r}'\right|\right)\phi_{F,\sigma}\left(\mathbf{r}\right)\phi_{F,\sigma}^{*}\left(\mathbf{r}'\right)\psi_{\sigma}\left(\mathbf{r}'\right)d\mathbf{r}'
\end{equation}
Next, we address the convolution in the random part of the above expression,
by rewriting the range-separated Coulomb potential as 
\begin{equation}
u_{C}^{\gamma}\left(\left|\mathbf{r}-\mathbf{r}'\right|\right)=\left\langle \zeta\left(\mathbf{r}\right)\zeta^{*}\left(\mathbf{r}'\right)\right\rangle ,\label{eq:u-stoch}
\end{equation}
where $\zeta\left(\mathbf{r}\right)=\left(2\pi\right)^{-3}\int d\mathbf{k}\sqrt{\tilde{u}_{C}^{\gamma}\left(\mathbf{k}\right)}e^{i\varphi\left(\mathbf{k}\right)}e^{i\mathbf{k}\cdot\mathbf{r}}$,
$\tilde{u}_{C}^{\gamma}\left(\mathbf{\mathbf{k}}\right)$ is the Fourier
transform of $u_{C}^{\gamma}\left(\mathbf{r}\right)$, and $\varphi\left(\mathbf{k}\right)$
is a random phase between $0$ and $2\pi$ at each $k$--space grid
point. This can be seen by inserting the definition of $\zeta\left(\mathbf{r}\right)$
into Eq.\ \eqref{eq:u-stoch} and using the identity $\left\langle e^{-i\left[\varphi\left(\mathbf{k}\right)-\varphi\left(\mathbf{k}'\right)\right]}\right\rangle _{\varphi}=\left(2\pi\right)^{3}\delta\left(\mathbf{k}-\mathbf{k}'\right)$.
(See Appendix~\ref{sec:The-k=00003D0-term} for the treatment of
the $k=0$ term). The non-local exchange operation is finally written
as:

\begin{align*}
\hat{k}_{\sigma}^{\gamma}\psi_{\sigma}\left(\mathbf{r}\right)= & -\left\langle \eta_{\sigma}\left(\mathbf{r}\right)\zeta\left(\mathbf{r}\right)\int\zeta^{*}\left(\mathbf{r}'\right)\eta_{\sigma}^{*}\left(\mathbf{r}'\right)\psi_{\sigma}\left(\mathbf{r}'\right)d\mathbf{r}'\right\rangle _{\xi,\varphi}
\end{align*}
\begin{equation}
+c\int u_{C}^{\gamma}\left(\left|\mathbf{r}-\mathbf{r}'\right|\right)\phi_{A,\sigma}\left(\mathbf{r}\right)\phi_{A,\sigma}^{*}\left(\mathbf{r}'\right)\psi_{\sigma}\left(\mathbf{r}'\right)d\mathbf{r}'.
\end{equation}
In actual applications we use a finite number $N_{\chi}$ of pairs
of stochastic orbitals $\chi_{\sigma}\left(\mathbf{r}\right)=\zeta\left(\mathbf{r}\right)\eta_{\sigma}\left(\mathbf{r}\right)$
and thus:
\begin{eqnarray}
\hat{k}_{\sigma}^{\gamma}\psi_{\sigma}\left(\mathbf{r}\right) & = & -\frac{1}{N_{\chi}}\sum_{\chi}\chi_{\sigma}\left(\mathbf{r}\right)\left\langle \chi_{\sigma}|\psi_{\sigma}\right\rangle \\
 & + & c\delta_{\sigma,\uparrow}\int u_{C}^{\gamma}\left(\left|\mathbf{r}-\mathbf{r}'\right|\right)\phi_{A,\sigma}\left(\mathbf{r}\right)\phi_{A,\sigma}^{*}\left(\mathbf{r}'\right)\psi_{\sigma}\left(\mathbf{r}'\right)d\mathbf{r}'.\nonumber 
\end{eqnarray}
The $\zeta\left(\mathbf{r}\right)$'s are calculated once and stored
in memory while the $\eta_{\sigma}\left(\mathbf{r}\right)$'s are
generated on the fly. The computational scaling of the non-local exchange
operation on $\psi_{\sigma}\left(\mathbf{r}\right)$ is thus $N_{\chi}N_{g}$
(vs. $10N_{g}log_{2}N_{g}\times N_{e}$ for the deterministic case).
Typically, $N_{\chi}=200$ and $N_{g}=10^{6}$ and thus, the operation
of the stochastic non-local exchange becomes comparable in terms of
computational effort to that of operating with the kinetic energy,
so the computational cost of applying the GKS Hamiltonian is similar
to that of the KS Hamiltonian.

\section{Results for Silicon Nanocrystals}

The new method has been implemented using the BNL functional\ \cite{Baer2005a,Livshits2007}
for a series of hydrogen passivated silicon nanocrystals of varying
sizes: $\mbox{\textrm{{Si}}\ensuremath{_{35}\textrm{{H}}_{36}}}$,
$\textrm{{Si}}\ensuremath{_{87}\textrm{{H}}_{76}},$ $\mbox{\textrm{{Si}}\ensuremath{_{147}\textrm{{H}}_{100}}}$,
$\mbox{\textrm{{Si}}\ensuremath{_{353}\textrm{{H}}_{196}}}$ and $\mbox{\textrm{{Si}}\ensuremath{_{705}\textrm{{H}}_{300}}}$
with real-space grids of $60^{3}$, $64^{3}$, $70^{3}$, $90^{3}$
and $108^{3}$ grid-points, respectively. We solve the generalized
KS equations fully self-consistently using the Chebyshev--filtered
subspace acceleration\ \cite{Zhou2006,Khoo2010} to obtain the occupied
and low lying unoccupied eigenfunctions and eigenvalues.

\begin{figure*}[!t]
\begin{centering}
\includegraphics[height=6cm]{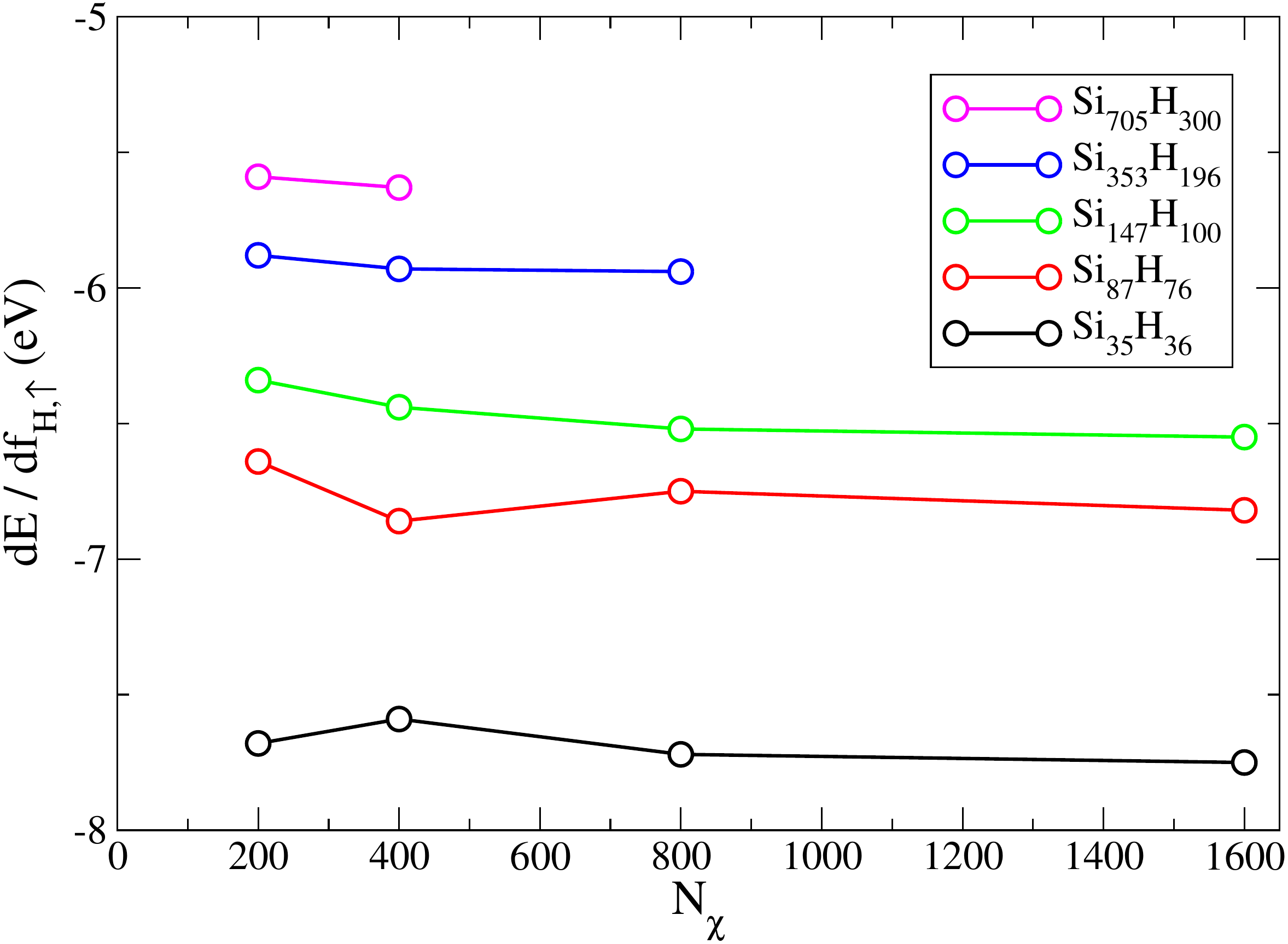}\ \ \includegraphics[height=6cm]{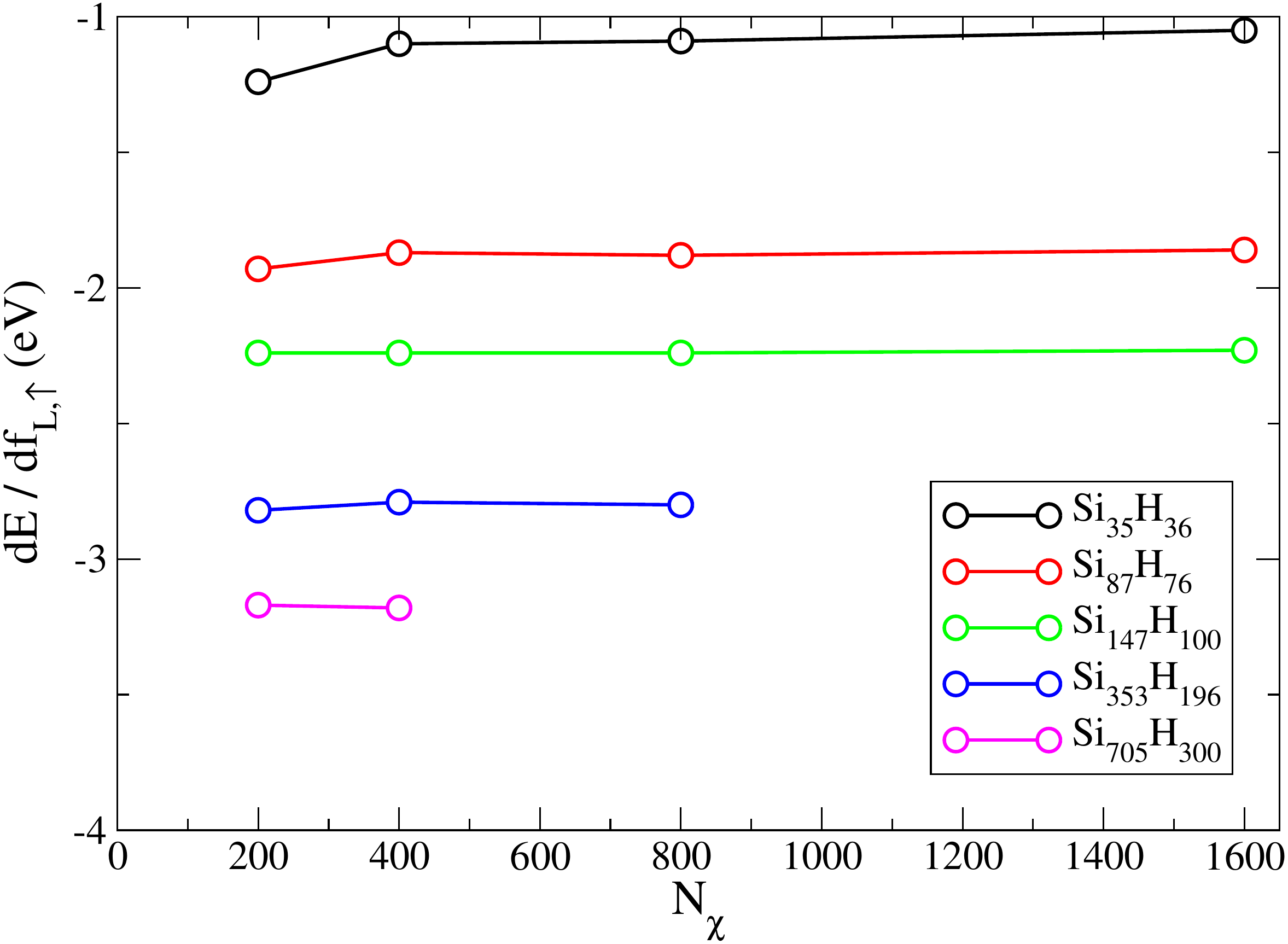}
\par\end{centering}

\protect\caption{\label{fig:Nchi} Convergence of the HOMO ($H$) energy (left panel)
and the LUMO ($L=H+1$) energy (right panel) with number of stochastic
orbitals $N_{\chi}$ for silicon nanocrystals using the BNL range-separated
functional. }
\end{figure*}

The curvature for the different NCs, estimated from a forward difference
formula $-\frac{\partial\varepsilon_{H,\uparrow}(c)}{\partial c}\approx\frac{\varepsilon_{H,\uparrow}\left(0\right)-\varepsilon_{H,\uparrow}\left(\delta\right)}{\delta}$
with $\delta=0.125$, is plotted as a function of $\gamma$ in Fig.\ \ref{fig:gamma}.
The curvature is a decreasing function of $\gamma$ and has a node
at the optimal value of the range parameter $\gamma_{*}$. For each
NC the curvature results are shown for several values of the number
of stochastic orbitals $N_{\chi}$. We find that the statistical fluctuations
near $\gamma_{*}$ become smaller as the system grows and can be reduced
with proper choice of $N_{\chi}$. For example, for the larger system
the results near $\gamma_{*}$ can be converged with only $N_{\chi}\approx200$
compared to the total number of occupied states for this system which
is $1560$. The reduction of these fluctuations is partially due to
the decrease of $\gamma_{*}$ itself as the NC size increases (this
decrease is shown in the right panel of Fig.\ \ref{fig:gamma}),
leading to a smaller contribution of the non-local exchange to the
orbital energies.

The results in the right panel of the figure also show that $\gamma_{*}$
closely follows a linear function of $N_{\mbox{Si}}^{-1/3}$. We expect
that for larger NCs with $N_{\mbox{Si}}>2500$, this linear relation
will break down and the optimal range parameter will converge to the
bulk value, which through reverse engineering\ \cite{Eisenberg2009}
can be estimated as $\gamma_{*}^{\infty}=0.02a_{0}^{-1}$ (shown as
a horizontal dotted line). Such a localization-induced by the exchange
has been seen for 1D conjugated polymers~\cite{Vlcek2015a} but not
for bulk solids like silicon, likely due to the enormity of the calculation.

In Fig.\ \ref{fig:Nchi} we plot the highest occupied molecular orbital
(HOMO, left panel) and lowest unoccupied molecular orbital (LUMO,
right panel) energies obtained from the relations~\cite{Janak1978}
\begin{eqnarray}
\varepsilon_{H,\uparrow} & = & -\left.\frac{\partial E_{RSH}^{\gamma}\left[\rho_{\uparrow},\rho_{\downarrow}\right]}{\partial c}\right|_{c\to0^{+}}\nonumber \\
\varepsilon_{L,\uparrow} & = & -\left.\frac{\partial E_{RSH}^{\gamma}\left[\rho_{\uparrow},\rho_{\downarrow}\right]}{\partial c}\right|_{c\to0^{-}}\label{eq:orbital-energies}
\end{eqnarray}
respectively, as a function of $N_{\chi}$ at $\gamma_{*}$. We find
that determining the HOMO and LUMO energies using the above first
derivative relations reduces the noise compared to obtaining their
values directly from the eigenvalues. Clearly $\varepsilon_{H,\uparrow}$
and $\varepsilon_{L,\uparrow}$ converge as $N_{\chi}$increases.
Moreover, as the system size increases the fluctuations in $\varepsilon_{H,\uparrow}$
and $\varepsilon_{L,\uparrow}$ decreases for a given value of $N_{\chi}$,
consistent with the discussion above. This is evident from the plot
of the differences between the frontier orbital energies at adjacent
values of $N_{\chi}$.

\begin{figure}[t]
\begin{centering}
=\includegraphics[width=8cm]{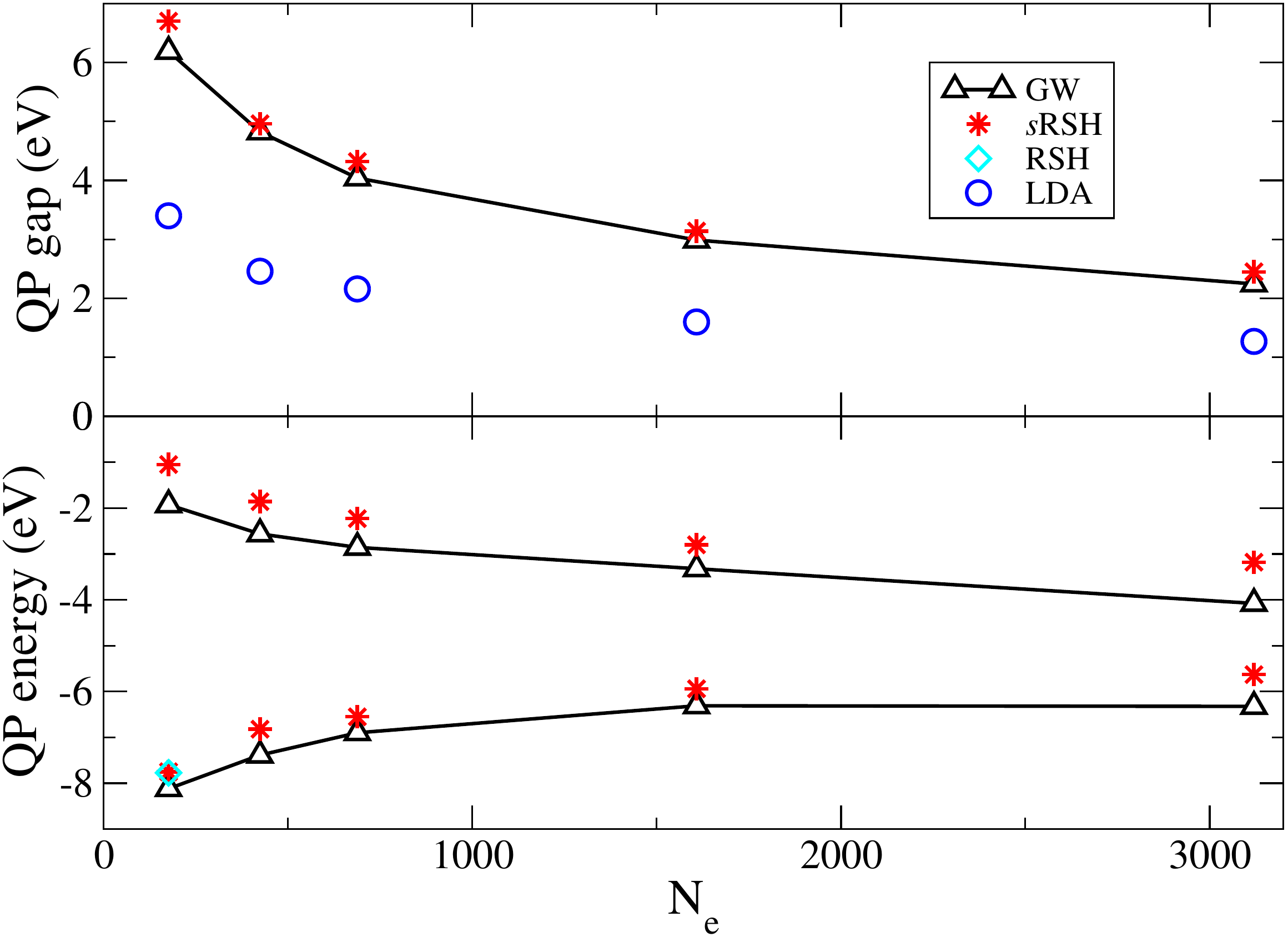}
\par\end{centering}

\protect\caption{\label{fig:qp} Lower panel: Comparison of the HOMO and LUMO energies
obtained using the \emph{s}GW approach (black triangles) and the stochastic
RSH within the BNL functional (red asterisk) for a series of silicon
nanocrystals. The cyan diamond represent the deterministic RSH within
the BNL functional. Upper panel: The corresponding quasiparticle band
gaps. Also shown is the DFT result within the LDA (blue circles). }
\end{figure}

In the lower panel of Fig.\ \ref{fig:qp} we plot the converged (with
respect to $N_{\chi}$) HOMO and LUMO energies at $\gamma_{*}$ for
the series of silicon NCs. For the smallest system ($\mbox{Si}_{35}\mbox{H}_{36}$)
we compare the stochastic approach developed here with a deterministic
RSH calculation using a non-local exchange with all occupied orbitals
and obtaining the Coulomb convolution integrals with FFTs, thereby
eliminating any source of statistical error. The purpose it to show
that when the stochastic results are converged the agreement with
a deterministic calculations is perfect on a relevant magnitude of
energy. We find that the HOMO energy increases and the LUMO energy
decreases with the size of the NC. This is consistent with our recent
calculations on silicon NCs using the stochastic GW approach, albeit
the fact that there is a small shift in the quasi-particle energies
obtained from the stochastic RSH approach compared to the \emph{s}GW.
Indeed, a similar shift has been reported previously for much smaller
silicon NCs.\cite{Stein2010} However, the source of this discrepancy
is not clear, particularly, since the GW calculations were done within
the so called \emph{$\mbox{\ensuremath{G_{0}}}\mbox{\ensuremath{W_{0}}}$}
limit, and the OT-RSH often provides better quasiparticle energies
in comparison to experiments.\cite{Kronik2012} In the upper panel
of Fig.\ \ref{fig:qp} we plot the fundamental (quasiparticle) gaps.
Here, the agreement with the \emph{s}GW approach is rather remarkable,
especially compared to the LDA results which significantly underestimate
the quasiparticle gap across all sizes studied. 

\begin{table}[b]
\begin{tabular}{|c|c|c|c|c|c|c|}
\hline 
System & Functional & $N_{\chi}$ & $\varepsilon_{H}\left(\mbox{eV}\right)$ & $\varepsilon_{L}\left(\mbox{eV}\right)$ & $\varepsilon_{g}\left(\mbox{eV}\right)$ & $T_{CPU}^{\left(c\right)}$\tabularnewline
\hline 
\hline 
\multirow{3}{*}{$\mbox{S\ensuremath{i_{35}H_{36}}}$} & LDA & - & -6.13 & -2.73 & 3.40 & 1.6\tabularnewline
\cline{2-7} 
 & \multirow{2}{*}{$\mbox{BN\ensuremath{L^{\left(a\right)}}}$} & 800 & -7.72 & -1.09 & 6.63 & 16\tabularnewline
\cline{3-7} 
 &  & 1600 & -7.75 & -1.05 & 6.70 & 30\tabularnewline
\hline 
\multirow{3}{*}{$\mbox{S\ensuremath{i_{705}H_{300}}}$} & LDA & - & -5.13 & -3.85 & 1.28 & 132\tabularnewline
\cline{2-7} 
 & \multirow{2}{*}{$\mbox{BN\ensuremath{L^{\left(b\right)}}}$} & 200 & -5.59 & -3.18 & 2.41 & 234\tabularnewline
\cline{3-7} 
 &  & 400 & -5.63 & -3.17 & 2.46 & 310\tabularnewline
\hline 
\end{tabular}

\begin{raggedright}
(a) $\gamma_{*}=0.148\, a_{0}^{-1}$
\par\end{raggedright}

\begin{raggedright}
(b) $\gamma_{*}=0.047\, a_{0}^{-1}$
\par\end{raggedright}

\begin{raggedright}
(c) In CPU-hrs
\par\end{raggedright}

\raggedright{}\protect\caption{\label{tab:Convergence-with-}Optimally-tuned BNL frontier orbital
energies and computational times $T_{CPU}$ vs. number of stochastic
orbitals $N_{\chi}$ for two (medium and large) silicon clusters.
Values for LDA are also given for comparison. As the system size grows,
$T_{CPU}$ for the optimally-tuned BNL decreases relative to the LDA
timings due to decrease of $\gamma_{*}$. }
\end{table}

In Table~\ref{tab:Convergence-with-} we provide numerical details
of the calculations for the smallest and largest NC studied. We report
the results for the HOMO and LUMO orbital energies for two different
choices of $N_{\chi}.$ Comparing these two values we can conclude
that the statistical errors for the LUMO are very small ($\approx0.01$~eV)
for the largest NC and even the HOMO has small errors of around $\approx0.05$~eV.
Moreover, similar or even larger statistical errors are observed for
the smaller NC for much larger values of $N_{\chi}$, indicating that
for a given accuracy the number of stochastic orbitals decreases with
the system size. This is partially correlated with the reduction of
$\gamma_{*}$ with the system size, as discussed above.

\section{Summary}

We have developed a stochastic representation for the non-local exchange
operator in order to combine real-space/plane-waves methods with optimally-tuned
range-separated hybrid functionals within the generalized Kohn-Sham
scheme. Our formalism uses two principles, one is a stochastic decomposition
the Coulomb convolution integrals and the other is the representation
of the density matrix using stochastic orbitals. Combining these two
ideas leads to a significant reduction of the computational effort
and, for the systems studied in this work, to a reduction of the computational
scaling of the non-local exchange operator, at the price of introducing
a statistical error. The statistical error is controlled by increasing
the number of stochastic orbitals and is also found to reduce with
the system size Applications to silicon NCs of varying sizes show
relatively good agreement for the band-edge quasiparticle excitations
in comparison to a many-body perturbation approach within the \emph{s}GW
approximation and excellent agreement for the fundamental band gap.
The stochastic approach has a major advantage over the $s$GW by providing
a self-consistent Hamiltonian which is central for post-processing,
for example in conjunction with a real-time Bethe--Salpeter approach.\cite{Rabani2015}
The results shown here for $N_{e}>3000$ and $N_{g}>10^{6}$ are the
largest reported so far for the optimally-tuned range-separated generalized
Kohn-Sham approach.

\begin{acknowledgments}

\end{acknowledgments}

R. B. and E. R. gratefully thank the Israel Science Foundation--FIRST
Program (Grant No. 1700/14). R.B. gratefully acknowledges support
for his sabbatical visit by the Pitzer Center and the Kavli Institute
of the University of California, Berkeley. D. N. and E. R. acknowledge
support by the NSF, grants CHE-1112500 and CHE-1465064, respectively. 

\appendix

\section{\label{sec:The-k=00003D0-term}treatment of the $k=0$ term}

For accelerating convergence, it turns out to be better to remove
the $\tilde{u}_{C}^{\gamma}\left(\mathbf{k=0}\right)$ term from the
the random vector expression representing the interaction, i.e., 
\[
\zeta\left(\mathbf{r}\right)=\left(2\pi\right)^{-3}d\mathbf{k}\sum_{\mathbf{k\ne}\mathbf{0}}\sqrt{\tilde{u}_{C}^{\gamma}\left(\mathbf{k}\right)}e^{i\varphi\left(\mathbf{k}\right)}e^{i\mathbf{k}\cdot\mathbf{r}}.
\]
This is because in practice the $\tilde{u}_{C}^{\gamma}\left(\mathbf{k}=\mathbf{0}\right)$term
is very large. Analytically, this term is easily shown to commute
with the Fock Hamiltonian and simply contribute a constant (times
the occupation) to the eigenvalues and to the total energy, so it
can be added a-posteriori: 
\[
\hat{k}_{\sigma}^{\gamma}\phi_{j,\sigma}\left(\mathbf{r}\right)\to\hat{k}_{\sigma}^{\gamma}\phi_{j,\sigma}\left(\mathbf{r}\right)-f_{j,\sigma}v_{0X},
\]

\[
\varepsilon_{j,\sigma}^{\gamma}\to\varepsilon_{j,\sigma}^{\gamma}-f_{j,\sigma}v_{0X},
\]

\[
K\left[\rho_{\uparrow},\rho_{\downarrow}\right]\to K\left[\rho_{\uparrow},\rho_{\downarrow}\right]-\frac{1}{2}v_{0X}\sum f_{j\sigma}^{2},
\]
where
\[
v_{0X}\equiv\left(2\pi\right)^{-3}d\mathbf{k}\tilde{u}_{C}^{\gamma}\left(\mathbf{k}=\mathbf{0}\right).
\]

\end{document}